\documentclass[apj]{emulateapj}

\usepackage{natbib}
\usepackage{amsmath}
\usepackage[colorlinks, citecolor=blue]{hyperref}
\shorttitle{Non-stellar Near-IR Emission in Galaxies at \MakeLowercase{z} $\sim 1$}
\shortauthors{Lange et al.}

\begin{document}
\title{Evidence for non-stellar rest-frame near-IR Emission associated with increased Star Formation in Galaxies at \MakeLowercase{z} $\sim 1$}

\author{Johannes U. Lange\altaffilmark{1}, Pieter G. van Dokkum\altaffilmark{1}, Ivelina G. Momcheva\altaffilmark{1}, Erica J. Nelson\altaffilmark{1}, Joel Leja\altaffilmark{1}, Gabriel Brammer\altaffilmark{2}, Katherine E. Whitaker\altaffilmark{3, 5}, Marijn Franx\altaffilmark{4}}

\altaffiltext{1}{Department of Astronomy, Yale University, New Haven, CT 06520, USA}
\altaffiltext{2}{Space Telescope Science Institute, 3700 San Martin Drive, Baltimore, MD 21218, USA}
\altaffiltext{3}{Department of Astronomy, University of Massachusetts, Amherst, MA 01003, USA}
\altaffiltext{4}{Leiden Observatory, Leiden University, Leiden, The Netherlands}
\altaffiltext{5}{Hubble Fellow}

\begin{abstract}
We explore the presence of non-stellar rest-frame near-IR ($2-5 \ \mu \mathrm{m}$) emission in galaxies at $z \sim 1$. Previous studies identified this excess in relatively small samples and suggested that such non-stellar emission, which could be linked to the $3.3 \ \mu \mathrm{m}$ polycyclic aromatic hydrocarbons feature or hot dust emission, is associated with an increased star formation rate (SFR). In this Letter, we confirm and quantify the presence of an IR excess in a significant fraction of galaxies in the 3D-HST GOODS catalogs. By constructing a matched sample of galaxies with and without strong non-stellar near-IR emission, we find that galaxies with such emission are predominantly star-forming galaxies. Moreover, star-forming galaxies with an excess show increased mid- and far-IR and H$\alpha$ emission compared to other star-forming galaxies without. While galaxies with a near-IR excess show a larger fraction of individually detected X-ray active galactic nuclei (AGNs), an X-ray stacking analysis, together with the IR-colors and H$\alpha$ profiles, shows that AGNs are unlikely to be the dominant source of the excess in the majority of galaxies. Our results suggest that non-stellar near-IR emission is linked to increased SFRs and is ubiquitous among star-forming galaxies. As such, the near-IR emission might be a powerful tool to measure SFRs in the era of the \textit{James Webb Space Telescope}.
\end{abstract}


\section{Introduction}

The near-infrared (NIR) emission of galaxies, corresponding to $\lambda_{\mathrm{rest}} \sim 2 - 5 \ \mu \mathrm{m}$, is becoming increasingly accessible with future facilities such as the \textit{James Webb Space Telescope} (JWST). However, the NIR spectral energy distribution (SED) of galaxies at $z \sim 1$ is still poorly understood. At rest-frame wavelengths $< 1 \ \mu \mathrm{m}$ the SED is dominated by stellar emission, while at wavelengths $> 10 \ \mu \mathrm{m}$ it is dominated by polycyclic aromatic hydrocarbons (PAH) and warm dust emission ($T \sim 100 \ \mathrm{K}$). It is the intermediate wavelength regime that has contributions from stellar emission, but possibly also other components.

Strong deviations from a purely stellar NIR SED have been used to identify active galactic nucleus (AGN) candidates \citep[see, e.g.][]{LacyWedge1, LacyWedge2, SternWedge, Spitzer-AGN, AGN-IRAC}. However, it is unlikely that AGNs and stars are the only sources of NIR emission \citep[see, e.g.][]{NIRE0, NIRE1, NIRE2, NIRE3}. \cite{NIRE1} found a systematic NIR excess with respect to the stellar emission in a sample of $88$ galaxies at $z \sim 1$ in the Gemini Deep Deep Survey. The results suggested that this emission is directly linked to the formation of stars, possibly originating from hot dust in circumstellar disks. This hypothesis was further corroborated by a study of $68$ nearby spatially resolved galaxies in the SINGS survey \citep{NIRE2}.

Given that the amount of star formation \citep{SFMS} as well as the NIR excess in galaxies in general \citep{NIRE3} increases when going to higher redshifts, it becomes evident that modeling the properties of high-redshift galaxies requires a better understanding of their NIR emission.

In this paper, we aim to take a further step toward understanding NIR emission in high-redshift galaxies through a multi-wavelength study in the GOODS fields, which offer deep ground- and space-based coverage, including \textit{Herschel} and \textit{Chandra} observations. We construct samples of galaxies with strong NIR excess emission and those lacking a detectable excess component, and we study their systematic differences. In the following, we assume a $\Lambda$CDM cosmology with $H_0 = 70 \ \mathrm{km} \ \mathrm{s}^{-1} \ \mathrm{Mpc}^{-1}$, $\Omega_{\Lambda} = 0.7$, and $\Omega_m = 0.3$.

\section{Observational Data}

\subsection{General Photometry}

We use the photometric catalogs from the 3D-\textit{HST} survey \citep{3D-HST}, as described in \cite{3D-HST_Photometry}. The photometric catalogs combine Wide Field Camera 3 (WFC3) imaging from the Cosmic Assembly Near-infrared Deep Extragalactic Legacy Survey \citep[CANDELS;][]{CANDELS1, CANDELS2} with publicly available data sets, including \textit{Spitzer}/IRAC data at $3.6 - 8 \ \mu \mathrm{m}$.

The IRAC photometry is of special interest to us since its four bands cover the rest-frame wavelength range $2 - 5 \ \mu \mathrm{m}$ at $z \sim 1$. Source flux densities were obtained inside a $3''$ diameter aperture due to the large point-spread function at wavelengths detected by the IRAC instrument. As described in detail in \cite{3D-HST_Photometry}, using the MOPHONGO code \citep{MOPHONGO}, high-resolution F125W, F140W, and F160W images were used as a prior in the source extraction to estimate contributions from blended sources. To additionally avoid blending we excluded galaxies that have a neighboring source within $1.5''$ or a source with a $\geq 3$ times higher $8.0 \ \mu \mathrm{m}$ flux density within $3''$.

\subsection{Mid- and Far-infrared Photometry}

$24 \ \mu \mathrm{m}$ \textit{MIPS} photometry was presented in \cite{SFMS}. We use the same catalogs where the bolometric IR luminosity at $8 - 1000 \ \mu \mathrm{m}$, $L_{\rm IR}$, is derived from a luminosity independent conversion of the $24 \ \mu \mathrm{m}$ flux density. The star formation rate (SFR) of a galaxy can then be estimated via
\begin{equation}
\mathrm{SFR} \ [M_\odot \ \mathrm{yr}^{-1}] = 1.09 \times 10^{-10} \ (L_{\rm IR} + 2.2 L_{\rm UV}) \ [ L_\odot ]
\label{eq:SFR_UR+IR}
\end{equation}
\citep{SFMS}, where $L_{\mathrm{UV}}$ is the rest-frame luminosity at $1215 - 3000$ \AA.

We also cross-match our photometric catalogs with the publicly available GOODS Herschel catalog \citep{Herschel} by requiring one source in the 3D-HST catalog to be within $0.5''$.

\subsection{Grism Spectroscopy}

Slitless \textit{HST}/WFC3 G141 grism spectra were obtained as part of the 3D-\textit{HST} survey and cover $1.1 < \lambda < 1.7 \ \mu \mathrm{m}$. We use the 3D-\textit{HST} v4.1.5 data release \citep{3D-HST_Grism}. A modified version of EAZY \citep{EAZY} was used to simultaneously fit photometric and grism data and obtain grism redshifts. For the small fraction of objects lacking spectroscopic redshifts or grism exposure we use photometric redshifts as described in \cite{3D-HST_Photometry} and \cite{3D-HST_Grism}; see \cite{Phot-z} for a detailed discussion of the photometric redshift accuracy.

Stellar masses were obtained from the $0.3 - 8 \ \mu \mathrm{m}$ photometry using FAST \citep{FAST} with a \cite{Chabrier} IMF and the \cite{Calzetti} attenuation law. Note that the NIR SED is down-weighted when fitting for redshifts and stellar parameters by assigning large systematic errors \citep{EAZY, FAST}. Therefore, an NIR excess should not significantly bias these properties.

\subsection{Chandra X-Ray Observations}

Chandra Deep-field South (CDF-S) and Chandra Deep-field North (CDF-N) cover the GOODS CANDELS fields and are the deepest \textit{Chandra} surveys available. We use the Chandra 4Ms \citep{Chandra4Ms} and Chandra 2Ms source catalogs \citep{Chandra2Ms, Chandra2Ms_Association, Chandra2Ms_Classification} for sources individually detectable in X-rays. For individually undetected sources in GOODS-S we use a stacking analysis \citep{X-RayStacking, ObscuredAGN1, ObscuredAGN2} on the exposure and count maps presented in \cite{Chandra4Ms}.

\section{Sample Selection}

We include galaxies in our sample with $0.5 < z < 1.5$. The redshift range is chosen such that H$\alpha$ measurements are available for the majority of the sample. Since we are interested in the characterization of the NIR emission of a galaxy, we require a minimum signal-to-noise ratio (S/N) in the $8.0 \ \mu \mathrm{m}$ flux density. To not artificially bias our sample toward galaxies with an NIR excess, we require the {\em expected}, rather than the measured, $8 \ \mu \mathrm{m}$ S/N to exceed 5. The expected stellar $8.0 \ \mu \mathrm{m}$ flux density is derived from fitting the $0.3 - 3.6 \ \mu \mathrm{m}$ photometry with EAZY. When fitting the SEDs, we assume an additional $5\%$ systematic error for every measured flux density, independent of the wavelength.

\begin{figure*}
  \plotone{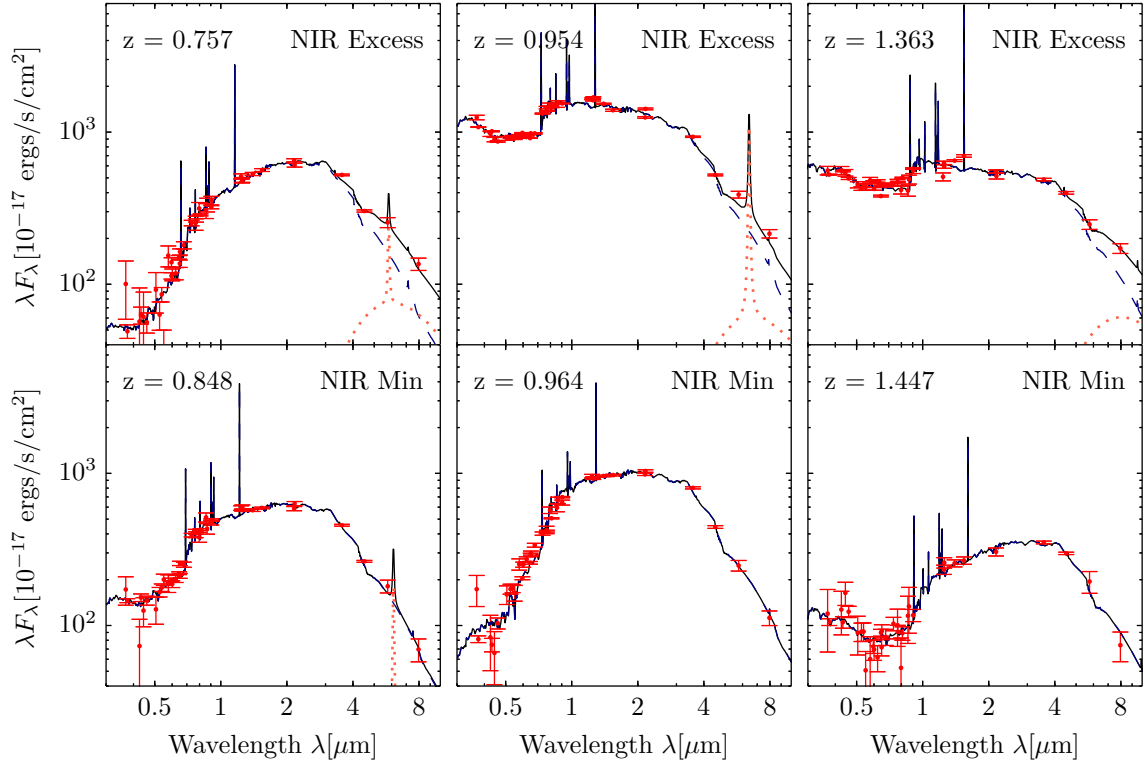}
  \caption{Example SEDs of galaxies in our sample. The top row shows NIR excess galaxies and the bottom row shows matched NIR min galaxies. The measured SED is shown by the red data points. The best-fit model is shown as a black solid line and composed of a stellar template (blue, dashed) and a PAH and graybody feature (red, dotted). The ratio of non-stellar to stellar $8.0 \ \mu \mathrm{m}$ flux density is $80\%$, $50\%$, and $50\%$ for the NIR excess galaxies from left to right.}
  \label{fig:SED}
\end{figure*}

We quantify the NIR excess emission by fitting the $0.3 - 8 \ \mu \mathrm{m}$ catalogs with a linear combination of the seven P\'{E}GASE models in EAZY, the $3.3 \ \mu \mathrm{m}$ PAH feature \citep{PAH} and an $850 \ \mathrm{K}$ graybody with a $\beta = 1$ emissivity \citep{NIRE1, NIRE3}. We note here that an NIR excess is for most galaxies only observable at $5.8$ and $8.0 \ \mu \mathrm{m}$, such that the combination of the PAH and the graybody feature can fit almost any observed excess. After fitting the UV to NIR SED, we require a good template fit, $\chi^2 / n_{\mathrm{filt}} < 3$. This results in a selection that is implicitly biased against galaxies whose overall SED is strongly dominated by an AGN, since we do not include AGN templates or account for time variability in our data set \citep[see, e.g.][]{AGN-Fitting}. We also exclude galaxies with bad GALFIT fits (GALFIT flag $>1$), signaling problems in the photometry, and strongly contaminated sources \citep{3D-HST_Photometry}. $1454$ galaxies pass the general selection cuts described thus far.

A visual inspection of the SEDs demonstrates that many galaxies have excess flux above the stellar SED in the long wavelength IRAC bands. We verified that the magnitude of this excess is too large to be caused by blending of neighboring sources in the majority of galaxies. We aim to elucidate the origin of this excess by selecting subsamples of galaxies with a strong NIR excess (hereafter NIR excess galaxies) and those lacking such a feature (hereafter NIR min galaxies). We define NIR excess galaxies as galaxies for which the $\chi^2$ of the EAZY fit, including the constant $5\%$ systematic error, reduces by $\Delta \chi^2 > 9$ when including the PAH and graybody feature, whereas NIR min galaxies have $\Delta \chi^2 < 1$. Six example galaxies are shown in Figure \ref{fig:SED}. We exclude from this analysis galaxies that fall into the intermediate range $1 \leq \Delta \chi^2 \leq 9$. It is worth pointing out that NIR min galaxies could still have a considerable NIR excess, in particular if the S/N in the IRAC bands is $\sim 5$. To allow an unbiased comparison between the two samples we construct pairs of galaxies from both groups matched by their redshift, stellar $8.0 \ \mu \mathrm{m}$ flux density, and $8.0 \ \mu \mathrm{m}$ exposure weight $w_{8.0}$. For this sample we require $\Delta z < 0.1$, $\Delta \log F_{8.0, \star} < 0.1$, and $\Delta \log w_{8.0} < 0.3$. As described in \cite{3D-HST_Photometry}, the weight is calculated from the inverse variance and accounts for background noise sources but not the Poisson noise due to sources themselves. Given that the NIR stellar light is a good proxy for the total stellar mass, we arrive at a roughly mass-matched sample. Additionally, pairs are chosen such that they have the same galaxy type, either star-forming or quiescent, based on the UVJ classification \citep{UVJ}. Altogether, we obtain a sample of $169$ galaxy pairs.

In the following, we will analyze these galaxy pairs to study systematic differences between the two groups.

\section{Results}

\subsection{UV to optical Photometry}

\begin{figure*}
  \plotone{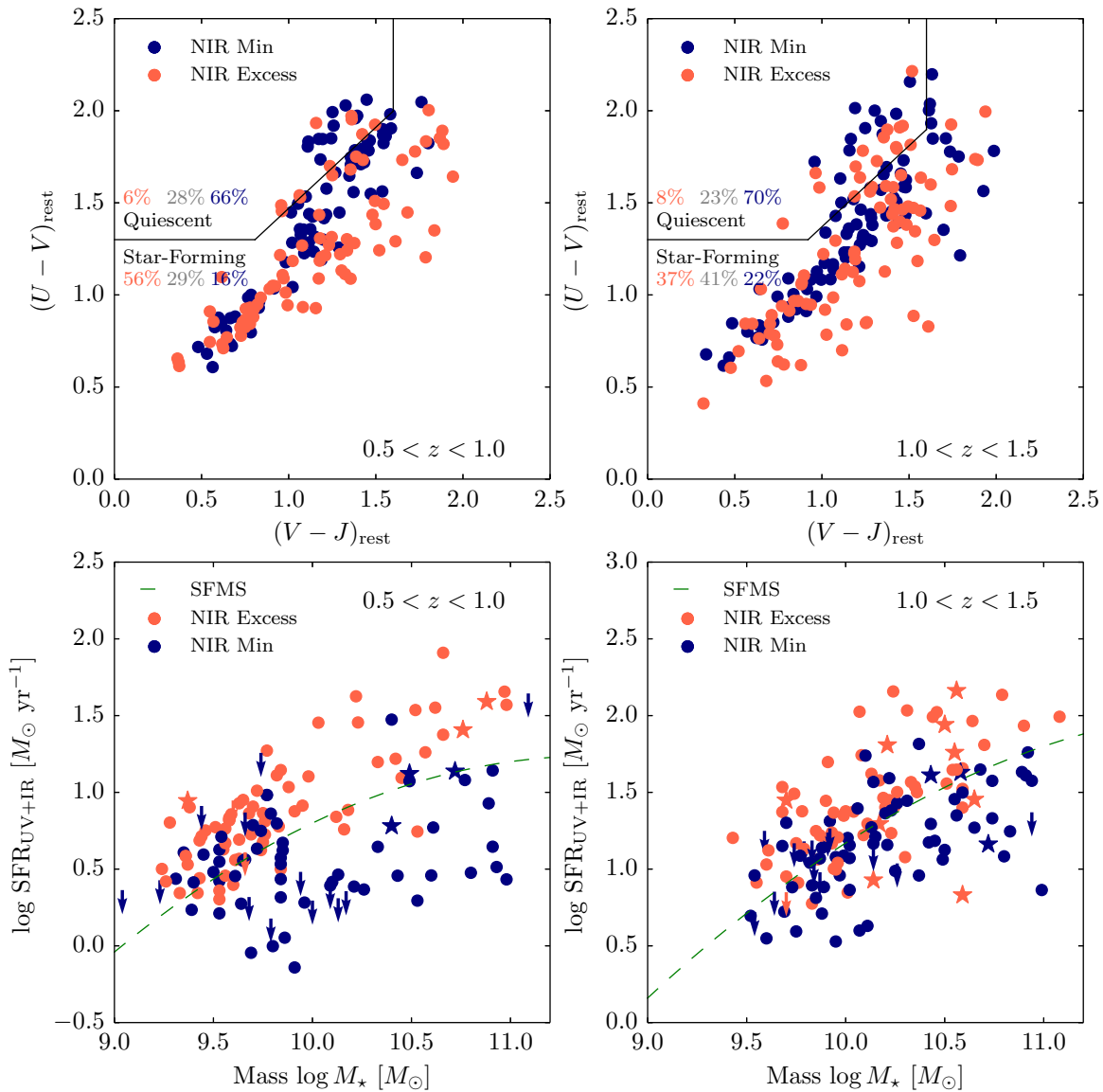}
  \caption{UVJ color diagrams and SFRs of matched galaxies in our sample. Blue dots show NIR min galaxies and red dots NIR excess galaxies. Top: the UVJ color--color space divided into star-forming galaxies and quiescent galaxies. We also show the fractions of star-forming and quiescent galaxies out of the unmatched sample of $1454$ galaxies that qualify as NIR excess or NIR min galaxies or belong to neither category (gray). Bottom: positions in the $\log(\rm SFR)$--$\log(M)$ plane for star-forming galaxies. Arrows denote $3\sigma$ upper limits for galaxies undetected at $24 \ \mu \mathrm{m}$ and stars denote X-ray detected AGN hosts. The dashed green line shows the polynomial fit of the SFMS presented in \cite{SFMS}.}
  \label{fig:UVJ_SFMS}
\end{figure*}

In the upper panels of Figure \ref{fig:UVJ_SFMS} we show the positions of galaxies in the UVJ diagram. Recall that we have selected NIR excess and NIR min galaxies based on their NIR emission, but not explicitly on their UV to optical spectrum. The most striking feature is that NIR excess galaxies tend to be star-forming galaxies, while NIR min galaxies are predominantly quiescent. We also note that among star-forming galaxies, NIR excess galaxies seem to occupy regions farther away from the quiescent clump than NIR min galaxies.

\subsection{Mid- and Far-infrared Emission}

We investigate where the star-forming galaxies fall compared to the star-forming main sequence (SFMS). This is shown in the bottom panel of Figure \ref{fig:UVJ_SFMS}, where the dashed green line is the fit presented in \cite{SFMS}. We see that NIR excess star-forming galaxies have noticeably higher SFRs than comparable NIR min galaxies. The median ratio of the IR luminosities derived from the $24 \ \mu \mathrm{m}$ flux densities is $\sim 2.5$ when comparing the NIR excess and NIR min galaxies. Interestingly, not only do NIR excess galaxies lie above the SFMS, but NIR min galaxies with $M_\star > 10^{10} \ M_\odot$, galaxies for which the upper limit on the non-stellar NIR contribution is very small, tend to fall below that relation. In other words, the absence of a strong NIR excess seems to indicate a reduced SFR.

While the $24 \ \mu \mathrm{m}$ flux density enhancements for NIR excess galaxies are compatible with increased SFR, they could also originate from AGN activity. Out of the sample of $169$ galaxy pairs, $61(41)$ NIR excess and only $14(12)$ NIR min galaxies have detections at $100(160) \mu \mathrm{m}$. The infrared colors of these galaxies can shed light onto what process drives the far-infrared emission of these galaxies. We find $F_{100} / F_{24} > 10$ ($F_{160} / F_{24} > 20$) for all but $10$ ($3$) NIR excess galaxies. This indicates that the FIR emission of these galaxies is dominated by star formation and not nuclear activity \citep[compare, e.g.][]{AGN-SFR5, AGN-SFR1}. Thus, the higher far-infrared detection rate of NIR excess galaxies further indicates an increased SFR compared to NIR min galaxies.

\subsection{\texorpdfstring{H$\alpha$}{H-alpha} Emission}

We compare the H$\alpha$ luminosities for pairs in which both galaxies have uncontaminated H$\alpha$ grism spectra and at least one is detected. Due to the low resolution of the grism spectra, H$\alpha$ and [\mbox{N\,{\sc ii}}] are blended. Using a sample of $41$ $(36)$ galaxy pairs, we find a median H$\alpha$ + [\mbox{N\,{\sc ii}}] luminosity ratio of $1.8$ in both GOODS-South and GOODS-North when comparing NIR excess to NIR min galaxies. Altogether, in $60$ of the $77$ pairs the NIR excess galaxy has a higher total measured H$\alpha$ + [\mbox{N\,{\sc ii}}] luminosity.

\begin{figure}
  \plotone{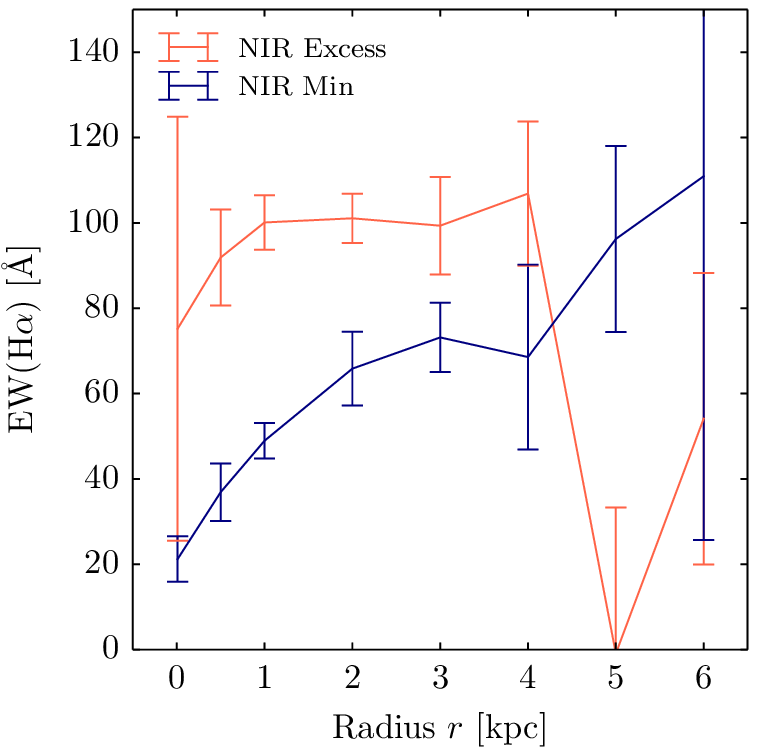}
  \caption{Radial EW(H$\alpha$) profiles for NIR excess and NIR min galaxies. Bands show the errors derived from bootstrap realizations.}
  \label{fig:StackedHalpha}
\end{figure}

Finally, we stack H$\alpha$ + [\mbox{N\,{\sc ii}}] maps for the two groups. The details of this procedure are described in \cite{HalphaStacks}. We only stack the emission of galaxies for which both members have uncontaminated H$\alpha$ grism spectra and do not host X-ray AGNs. As shown in Figure \ref{fig:StackedHalpha}, we find that the increase of the H$\alpha$ equivalent width of NIR excess galaxies is not limited to the center, as might have been expected if AGNs were the sole cause of the excess. Instead, the equivalent width is enhanced at least to $\sim 3 \ \mathrm{kpc}$, compared to NIR min galaxies, consistent with star formation being the dominant cause.

\subsection{X-Ray Emission}

We find that $26$ NIR excess and $10$ NIR min galaxies are detected in X-rays with \textit{Chandra}. $19$ of the NIR excess and $8$ of the NIR min X-ray sources have been identified as secure AGNs by \cite{Chandra4Ms} or \cite{Chandra2Ms_Classification}. Given this result, it is likely that at least in $5\%$ of the galaxies the NIR excess is caused by weak or obscured AGNs.

We test the presence of a large population of weak or obscured AGNs using a stacking analysis \citep{X-RayStacking} of galaxies in GOODS-South. Excluding galaxies that host confirmed X-ray AGNs themselves or are near detected X-ray sources, we stack the X-ray emission of $74$ NIR excess and $83$ NIR min sources. We compare this to the expected flux from star formation alone for which we use the relation from \cite{XRAY-SFR2} and assume an intrinsic power-law index of $\Gamma = 2.0$. We use the UV+IR derived SFRs and correct for galactic absorption in the Milky Way. For the relation by \cite{XRAY-SFR2} we take into account the intrinsic log-normal scatter and a correction factor of $1.4$ because of the systematically different SFRs derived in \cite{SFMS}.

We find confident detections for all source groups and bands. The soft- and hard-band fluxes for the NIR excess galaxies are $9.8 \times 10^{-9} \ \mathrm{cm}^{-2} \mathrm{s}^{-1}$ ($14.0\sigma$) and $3.0 \times 10^{-9} \ \mathrm{cm}^{-2} \mathrm{s}^{-1}$ ($3.2\sigma$). Similarly, the values for the NIR min galaxies are $4.8 \times 10^{-9} \ \mathrm{cm}^{-2} \mathrm{s}^{-1}$ ($7.2 \sigma$) and $2.7 \ \times 10^{-9} \mathrm{cm}^{-2} \mathrm{s}^{-1}$ ($3.0 \sigma$). The expected fluxes from SFR alone are $9.3$, $2.4$, $3.6$, and $0.9 \times 10^{-9} \ \mathrm{cm}^{-2} \mathrm{s}^{-1}$, respectively. Given the uncertainties and scatter in the SFR--$L_X$ relation, we find that the measured fluxes are broadly consistent with being primarily driven by star formation.

\subsection{SFR--NIR Excess Relation}

\begin{figure*}
  \plottwo{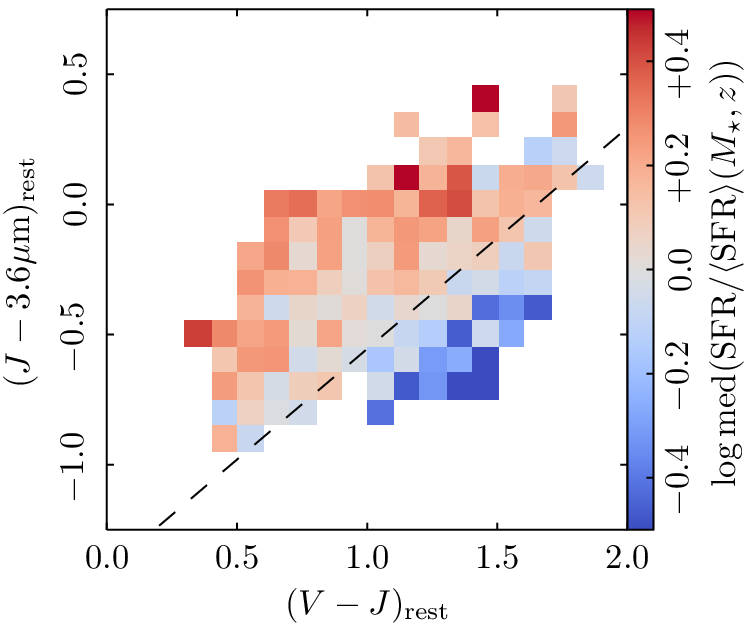}{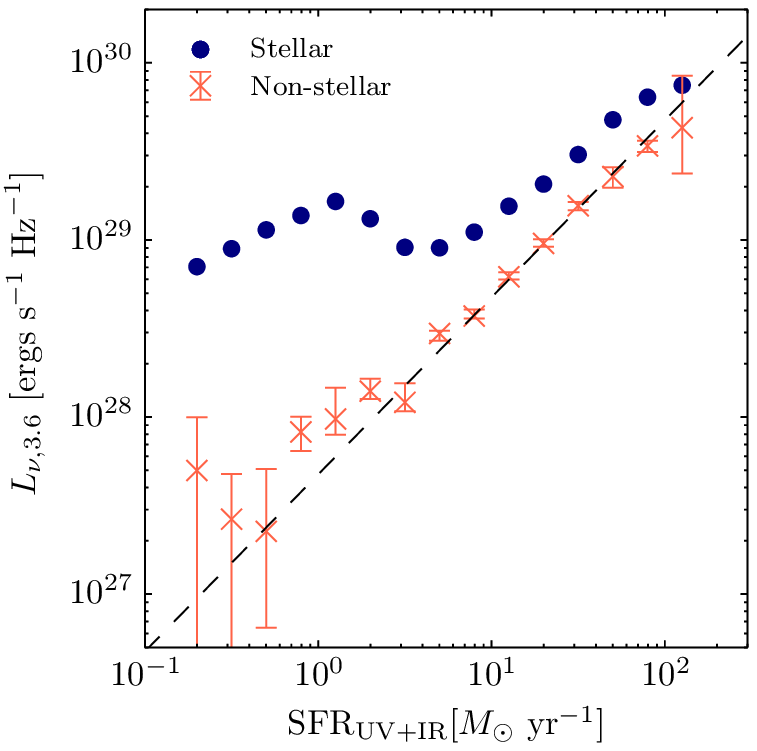}
  \caption{Left: rest-frame $V-J$ vs. $J - 3.6 \ \mu \mathrm{m}$ colors of all star-forming galaxies in our sample. Colors indicate the median star formation rate with respect to the SFMS. Redder NIR colors correlate strongly with increased SFRs. Right: median rest-frame IRAC1 stellar and non-stellar luminosity as a function of SFR for all galaxies in the broader sample, except X-ray AGNs. Error bars are computed from $100,000$ bootstrap realizations and the dashed line shows the best-fit linear relation.}
  \label{fig:3p6}
\end{figure*}

After examining qualitative trends in a matched sample of galaxies, we analyze the entire unmatched sample. The left panel of Figure \ref{fig:3p6} shows rest-frame $V-J$ vs. $J - 3.6 \ \mu \mathrm{m}$ colors of all star-forming galaxies in $0.5 < z < 1.5$. The rest-frame colors were derived from the best-fit EAZY model including the NIR templates. The dashed line effectively separates NIR excess from NIR min sources. $85\%$ of all NIR excess sources, regardless of having a matched counterpart, fall above the line, while $85\%$ of the NIR min galaxies fall below. We also show the median ratio of the SFR with respect to the relation in \cite{SFMS}. From this figure it is evident that an increased $J - 3.6 \ \mu \mathrm{m}$ color, representing an NIR excess, correlates well with increased $\mathrm{SFR}_{\rm UV+IR}$. In the right panel of Figure \ref{fig:3p6} we explicitly show the relation between the NIR excess, parameterized by the rest-frame $3.6 \ \mu \mathrm{m}$ luminosity in excess of the expected stellar emission, and the SFR. We find a linear relation of the form
\begin{equation}
\nu L_{\nu, 3.6}[\mathrm{erg} \ \mathrm{s}^{-1}] = (4.0 \pm 0.2) \times 10^{41} \times \mathrm{SFR}_{\rm UV + IR} [M_\odot \mathrm{yr}^{-1}].
\end{equation}

\section{Discussion}

In this Letter we quantify and explore the origin of the NIR excess over purely stellar light in galaxies at $z \sim 1$. We find that such an excess emission is particularly common among star-forming galaxies and that the strength of the excess scales linearly with $\mathrm{SFR}_{\rm UV+IR}$. We also explore the possibility that the excess could come from weak or obscured AGNs, but find this explanation for the majority of the galaxies unlikely based on their FIR emission and X-ray luminosities and the fact that the excess is prevalent in star-forming galaxies, while being almost absent in quiescent galaxies.

\begin{figure}
  \plotone{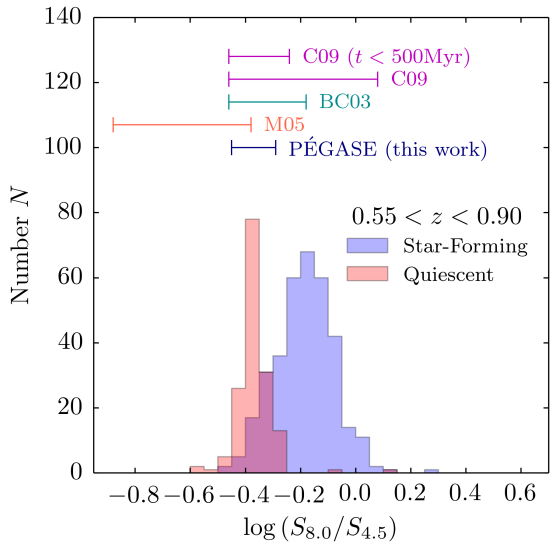}
  \caption{Distribution of $[8.0] - [4.5]$ IRAC colors for galaxies in our sample with $0.55 < z < 0.90$. We also show the ranges allowed by our P\'{E}GASE templates and other SPS models. Star-forming galaxies show systematically redder colors, indicative of excess emission redward of the $3.3 \mu \mathrm{m}$ PAH feature.}
  \label{fig:IRAC-Colors}
\end{figure}

A major uncertainty in the present study is the template library used. In particular, different SPS models predict subtly different SEDs in the NIR, primarily due to their different treatment of dust around AGB stars \citep{EzGal, AGB}. An analysis of all theoretical uncertainties in the models is beyond the scope of this Letter. We show in Figure \ref{fig:IRAC-Colors} the $[8.0] - [4.5]$ IRAC colors of $0.55 < z < 0.90$ galaxies without X-ray AGNs fulfilling our selection cuts. Using the stellar libraries compiled by \cite{EzGal} we show the predictions by \citet[][hereafter BC03]{BC03}, \citet[][M05]{M05} and \citet[][C09]{C09} for declining star formation histories with $\tau = 0.1 \ \mathrm{Gyr}$ for all ages, metallicities, and redshifts. We also include the range of the P\'{E}GASE templates used in the analysis. Only the C09 models span the observed range of IRAC colors, including star-forming galaxies. However, the very red NIR colors come from stellar populations older than $1 \ \mathrm{Gyr}$. If we limit the models to ages below $500 \ \mathrm{Myr}$, models with significant recent star formation, the maximum value reduces to $-0.24 \ \mathrm{dex}$ and lies below the median value for star-forming galaxies. In principle, dust extinction could also lead to redder NIR colors. However, explaining the difference of $0.2 \ \mathrm{dex}$ between star-forming and quiescent galaxies by optically thick star formation would require $A_V \sim 10$ or higher \citep[compare, e.g.][]{IR-Extinction_1, IR-Extinction_2} and is highly unlikely for the entire population of star-forming galaxies.

We conclude that an NIR excess seems to be a common property of star-forming galaxies and is correlated with increased SFR. Thus, the NIR regime, at the moment mostly neglected when fitting galaxy SEDs, offers additional information about star formation in high-$z$ galaxies. In the near future, spectroscopy with JWST/NIRSpec will elucidate the physical origin of the excess and can provide more information on the origin of the relation in Figure \ref{fig:3p6}.

\acknowledgements

We thank the CDF-S, CDF-N, and the GOODS-Herschel teams for making their reduced data publicly available. We are grateful to Charlie Conroy and Jonathan Trump for helpful comments and discussion. KEW gratefully acknowledges support by NASA through Hubble Fellowship grant \#HF2-51368 awarded by the Space Telescope Science Institute, which is operated by the Association of Universities for Research in Astronomy, Inc., for NASA, under contract NAS 5-26555. This work is based on observations taken by the 3D-HST Treasury Program (GO 12177 and 12328) with the NASA/ESA HST, which is operated by the Associations of Universities for Research in Astronomy, Inc., under NASA contract NAS5-26555.

\end{document}